\documentstyle[12pt]{article}
\begin{document}
\begin{center}
{\Large \bf Unified Compositeness\\
of Leptons, Quarks and Higgs Bosons}\footnote{Submitted 
to the 28th 
International Conference on High
            Energy Physics, 25-31 July 1996, Warsaw, Poland}  

\end{center}
\begin{center}
{\large Vasily V.\ Kabachenko}\footnote{  
Moscow Institute of Physics and Technology, Russia}
{\large and Yury F.\ Pirogov}\\
{\it Institute for High Energy Physics, Protvino, Russia}\\
\end{center}

{\small
The unified compositeness  of leptons, quarks and Higgs bosons
is proposed as a possible scenario for New Physics 
beyond the Standard Model. The following topics of the scenario 
are briefly discussed:
\newline $\bullet$
Chiral gauge exceptional symmetry $E_6$ as a strong internal binding mechanism;
\newline $\bullet$ Higgs doublet as a composite Goldstone boson; 
\newline $\bullet$
Nonlinear Standard Model as a prototype ``low energy'' effective field theory 
of the unified compositeness;
\newline $\bullet$
Hidden local symmetry and an improved ``low energy'' effective field theory 
of the unified compositeness;
\newline $\bullet$
Heavy composite vector bosons 
and vector boson dominance of the SM gauge interactions;
\newline $\bullet$
Universal dominant residual interactions as a signature of the 
unified compositeness;  
\newline $\bullet$
Manifestations of the residual interactions 
and potential of the future TeV $e^+e^-$ linear colliders to uncover the 
unified compositeness.}

\section*{Introduction}

Are leptons and quarks composite or not? This is the question.
The same is for Higgs bosons. If both these types of the Standard Model 
(SM) fields were composite simultaneously, having common 
substructure, one could, in principle, solve a lot of the SM problems.
First of all, considering leptons and quarks as light 
composite fermions one could find a rationale for the 
well-known generation problem and that of the fermion quantum 
numbers~(see, e.g.,~\cite{peskin}). 
Further, treating the SM Higgs doublet as composite Goldstone 
boson~\cite{georgi} one could solve the naturalness problem~\cite{thooft} 
of the SM without supersymmetry. More than that, one could also try to unify 
Higgs self-interactions and Yukawa interactions reducing thus 
their arbitrariness. This interactions should emerge as residual ones from an 
interplay of the hyperstrong binding gauge interactions and the perturbative 
weak gauge interactions, etc.  

In a series of papers~\cite{pirogov1}--\cite{pirogov3} one of the present 
authors (Yu.F.P.) has developed the scheme of the unified
compositeness  of leptons, quarks and Higgs bosons
(gauge bosons being still elementary)
as a promising scenario for New Physics beyond the SM.
Some phenomenological consequences of the scenario have been further studied
by us in a number of subsequent papers~\cite{kabachenko1,kabachenko2}. 
In this report we present a brief survey of these developments.

\section{Scenario of unified compositeness}

Let us expose the principle ideas of the scenario of unified compositeness of
leptons, quarks and Higgs bosons. The scenario encounters two stages: that
of the dynamical symmetry breaking and that of the spontaneous symmetry
breaking. 

\paragraph{Dynamical symmetry breaking}

Let a hypothetical hyperstrong gauge theory $S_{loc}$, 
responsible for the tight internal binding of the SM composite particles, 
possesses a global chiral symmetry $G$, that of the Lagrangian. As a result 
of the nontrivial topological structure in $S_{loc}$ at a scale $\cal F$
a partial dynamical symmetry breaking $G\to H$ takes place, where $H$
is a residual symmetry of the vacuum $|0\!>$. It is supposed that 
$H$ embeds the symmetry of the SM: $H\supseteq I_{SM}=SU(2)_L\times U(1)_Y$. 
In this, the broken symmetry $G/H$ corresponds to true 
Goldstone bosons, in particular, to the Higgs doublet. On the other hand, the 
unbroken chiral symmetry $H$ is responsible via the 't~Hooft anomaly matching 
condition for the appearance of the massless composite fermions in addition 
to the massless composite Higgs bosons. 

In reality, a part of the Lagrangian symmetry $G$ is gauge: $G\supset
I_{loc}\supset I_{loc}^{SM}$. Thus this symmetry undergoes the partial 
dynamical
braking as $I_{loc}\to R_{loc}$  with  the residual gauge symmetry being 
$R_{loc}= I_{loc}\cap H$  and $R_{loc}\supseteq  I_{loc}^{SM}$. 
The Goldstone bosons corresponding to the broken part of the local symmetry 
$I_{loc}/R_{loc}$ are absorbed via Higgs mechanism by the proper gauge 
bosons $V$, the latter becoming massive: $M_V\simeq g_V \cal F$. The rest of
the Goldstone bosons, including the Higgs doublet, is still massless at this
stage.  

In the framework of the effective field theory the dynamical symmetry 
breaking is
described by the nonlinear model $G/H$~\cite{weinberg1,callan} 
with intrinsic gauge theory $I_{loc}$ being spontaneously
broken as $I_{loc}\to R_{loc}$. Here the weak gauge 
interactions are considered as a perturbation not of importance in the lowest 
approximation for the basic properties of the symmetry breaking pattern. 
Still, interactions of $I_{loc}$ explicitly violate symmetry $G$, and their 
account results in important physical effects.

\paragraph{Spontaneous symmetry breaking}

At $I_{loc}$ being turned off, all the possible orientations of the residual
symmetry $H$ inside the total symmetry $G$ are equivalent. This results in a 
set of the degenerate $H$ invariant vacua $|\xi\!>\equiv\xi |0\!>$, where 
$\xi \in G/H$. Explicit violations due to $I_{loc}$ being turned on,
this equivalence is lost. The question arises as to what is the preferred
orientation of $H$ relative $I_{loc}$ and what is the true vacuum of the
lowest energy? This is the so-called vacuum alignment 
problem~\cite{dashen,peskin2}. 
The orientation in question is determined by the radiative corrections 
caused by the virtual emission and absorption of the $I_{loc}$ gauge bosons.
 
Namely, let us consider  the effective action 
$\Gamma(\phi)=-V(\phi)_{eff}+\dots$, where  the one-loop effective potential
is $V(\phi)_{eff}=\mu^2\phi^\dagger\phi+\lambda(\phi^\dagger\phi)^2+
{\cal O}((\phi^\dagger\phi)^3)$. The following general statement is
true~\cite{dashen,peskin2}. 
The radiative corrections due to the dynamically unbroken symmetry $R_{loc}$
contribute positively to $\mu^2$ and thus stabilize the unperturbed vacuum 
$|0\!>$ trying to orient the residual global symmetry $H$ along the symmetry 
$R_{loc}$ itself. And v.v., the radiative corrections due to dynamically 
broken symmetry $I_{loc}/R_{loc}$ contribute negatively to $\mu^2$ and hence 
destabilize  vacuum $|0\!>$ trying to disorient symmetry $H$ relative to 
symmetry $R_{loc}$. The net effect is $\mu^2=C{\bar g}^2/(4\pi)^2\,{\cal F}^2$,
where $\bar g$ is a generic effective gauge coupling constant, $\cal F$ is 
the mass scale of the dynamical symmetry breaking and $C$ is a numerical being 
determined by the ratio of the two contributions with opposite signs. 

Thus there are tree possibilities for the curvature $\mu^2$ of the 
effective potential.
\begin{itemize}
\item{$\bf\mu^2\ge 0$}
This is the case of the convex  potential. Here $R_{loc}$ is left unbroken, 
and Goldstone bosons $\phi$ turn into the pseudo-Goldstone ones with  
$m^2\ge 0$. 
\item{$\bf\mu^2 < 0$}
In this case the potential is concave. This means that the symmetry 
$R_{loc}$ is spontaneously broken.  Goldstone bosons $\phi$ turn into the 
would-be Goldstone ones and are absorbed through the Higgs mechanism by 
the corresponding weak gauge bosons. 
\item{$\bf\mu^2=0$}
This is degenerate case of the flat one-loop potential. Here one-loop
approximation is insufficient and two-loop corrections have to be
considered~\cite{cosower}.
\end{itemize}  

We take for granted that one of the two last cases is realized, and the
spontaneous symmetry breaking of $R_{loc}$ takes place. The Higgs
self-interactions described by $V_{eff}\simeq V_H$ are no longer fundamental
but arise as residual ones from more fundamental gauge interactions.
Similarly, the effective Yukawa interactions must arise as a result of the 
radiative effects due to $I_{loc}$. Both these types of interactions are 
deprived of their fundamental
status. Thus a kind of the unification of the Higgs
and Yukawa interactions occurs, and there appear an opportunity, at least in
principle, to reduce their arbitrariness.

Now, there are two possibilities for the mass scale $\cal F$ of the unified 
compositeness.
\begin{itemize}
\item{\bf TeV compositeness}
It realizes in the most general one-loop case. Here one can show that 
${\cal F}={\cal O}(v)$, where $v$ is the SM v.e.v. We consider this as
phenomenologically unacceptable. For ${\cal F}\gg v$ to take place a fine tuning
is required, and this is unnatural.

\item{\bf Deca-TeV compositeness}
For the theory to be natural, one should put to one-loop $v\equiv 0$.
Then in two-loops one has $\mu^2={\cal O}((g^2/(4\pi)^2)^2{\cal F}^2)$, whereas
$\lambda={\cal O}(g^2/(4\pi)^2)$ as before. It follows hereof that
$v\equiv\mu/\sqrt{\lambda}={\cal O}(g/4\pi\,\cal F)$, or 
$v={\cal O}(\sqrt{\alpha_W/4\pi}\cal F)$. In other words, one has 
${\cal F}={\cal O}(m_W/\alpha_W)$, or ${\cal F}={\cal O}(10\,{\mathrm TeV})$.
Thus, the natural two-loop hierarchy ${\cal F}\gg v$ 
between compositeness scale and Fermi scale arises.  
\end{itemize}
It is this last scenario that is developed in what follows. More details can be 
found in refs.~\cite{pirogov1}--\cite{pirogov3}.

\section{Chiral gauge exceptional symmetry}

A paramount problem in building a realistic composite model of 
leptons and quarks is to find underlying forces capable of binding 
these particles at the distances much smaller then their Compton 
wave lengths. Strongly coupled non-Abelian gauge theories $S_{loc}$ 
provide presently a unique well-fitted framework for such a binding mechanism.

It is imperative that in the process of confinement 
a set of (almost) massless composite fermions should emerge.  
In other words, this is to require that some residual chiral 
symmetry should be left unbroken in the transition. A necessary (but not 
sufficient) condition for this is the chiral anomaly 
matching condition~\cite{thooft,frishman}.

There are  conclusive  arguments  that  strongly  interacting 
$SU(N)$ gauge theories with $n$ Dirac constituent  fermions  (so  that 
their  $SU(N)$  representation  is  vector-like)  break  the  chiral 
symmetry $SU(n)\times SU(n)\times U(1)$ down to the vector-like  one  
$SU(n)\times U(1)$ and  hence  do  not  produce  massless  composite   
fermions~\cite{witten,weingarten}.      
Similarly, for confining groups $SO(N)$ ($Sp(N)$) with strictly 
real (resp.,  pseudo-real)  representations  the  chiral  symmetry 
$SU(n)$ of $n$ Weyl fermions is likely to be broken down to the vector 
one $SO(n)$ (resp., pseudo-vector one 
$Sp(n)$)~\cite{peskin2,dimopoulos,peskin3}. 
If so, the only candidates to be considered at all for composite model  
purposes are the  non-Abelian  gauge  symmetries with complex 
(non-self-contragradient) representations.

It is well-known that, restricting oneself by the simple  Lie 
groups, one encounters just three such possibilities: 
$SU(N),  N  \geq  3$; $SO(4k+2), k \geq  2$ 
and exceptional group $E_6$  (see,  e.g.,~\cite{okubo}). 
The complex representations of the $SU(N)$ group can be anomaly free 
only if they contain necessarily higher rank tensors (in line with 
the fundamental ones, if  desired).  The  $SO(4k+2)$  group,  though 
being anomaly free, does not admit composite fermions  built  only 
of the constituent fermions in the fundamental (even  dimensional  spinor) 
representations.

On the other hand, $E_6$  group is free from both these drawbacks. 
First  of  all,  $E_6$   is  anomaly  safe~\cite{okubo} in $d=4$ dimensions
so that there are no 
restrictions on its chiral fermion content. Besides, it possesses the 
odd (namely, the third) rank invariant tensor in  the  fundamental 
representation~\cite{cvitanovic} and hence could lead to the  required  composite 
fermions.

Therefore one concludes that if one sticks to fermions in the  fundamental 
representations  of simple Lie  groups, only chiral $E_6$ is 
permissible as $S_{loc}$. In this, the semi-simple Lie groups and/or 
nonfundamental  representations,  though  not  being  excluded  
a~priori, nevertheless seems quite unnatural for a truly  underlying 
theory one is searching for.

\paragraph{Partial chiral symmetry breaking}

A priori, for a strongly  interacting  gauge  theory $S_{loc}$  
with chiral fermions there are two alternatives: either gauge  symmetry 
is tumbled dynamically through its own strong  interactions  until 
all the constituent fermions  are  allowed  to  acquire  dynamical 
masses, or the gauge symmetry remains exact and some of the chiral 
fermions have to remain massless. (In principle,  some  intermediate 
patterns could be adopted too.) It is the result of  the  dynamical 
competition between chiral symmetry breaking and confinement: which 
of these  possibilities  will  win.  Presently  one  does not  know 
dynamical  conditions  under  which  either  of  them   could   be 
realized~\cite{peskin3}.

It is the second pattern (or at least  some  admixture  of  it) 
that is required for a composite picture of leptons  and  quarks  to 
have any dynamical reason at all. Therefore, we take for  granted 
that in the case under consideration underlying  strongly  coupled 
gauge symmetry $E_6$   is  preserved,  and  proceed with  studying  the 
ensuing pattern of chiral symmetry breaking.

So let for the chiral gauge theory $S_{loc}$, the Lagrangian chiral 
symmetry $G$ be dynamically broken to some vacuum residual  symmetry 
$H: G \to  H$. This is supposed to  take  place  due  to  formation  of 
vacuum  bilinear  condensate  
$<\chi_L \bar \chi_R >$  (plus  $<\chi_R \bar \chi_L>)$   from   the 
constituent  Weyl fermions $\chi_{L,R}$.  We  postulate  that  in   this 
transition all those and  only  those  constituents,  which  can  get 
massive without breaking the confining gauge symmetry,  do  acquire 
dynamical masses. More than that, these masses are  assumed  to 
be equal. In other words, the  hypothesis states that  the  residual 
chiral symmetry $H$ is the maximal one consistent with the dynamical 
mass  generation  and  preservation  of the strongly   coupled   gauge 
symmetry.  This  agrees  with  the  pattern  of  chiral   symmetry 
breakdown adopted for vector-like, vector and pseudo-vector  gauge 
theories, resp.\ $SU(N)$, $SO(N)$  and  
$Sp(N)$~\cite{witten}--\cite{dimopoulos}.  
Though this is just a  hypothesis (a kind of
the ``survival'' hypothesis) it is well-formulated  and is 
more  predictive than  mere  postulating  some  breaking pattern.

More explicitly, let in a general case of the chiral gauge 
$E_6$  symmetry  there  be  $l$  left-handed  and  $r$  right-handed  Weyl 
fermions $\chi_L$ and $\chi_R$ transforming as $E_6$ fundamental  
representation $\underline N = \underline {27}$. 
(Equivalently, in terms of left-handed fermions only,  let 
there be $l$ of $\underline N$'s and $r$ of $\overline{N}$'s.) 
In general, $l \neq  r$ are  arbitrary. 
Asymptotic freedom of the gauge $E_6$  requires only that 
$(l + r)  < 22$. 
For definiteness let us assume that $l \geq  r \geq  0$. The Lagrangian 
chiral symmetry $G$, left unbroken by the $E_6$ instantons, looks  at 
different $r$ as follows $(l\geq 2)$:
\begin{equation}
G=\left \{ 
\begin{array}{l}
SU(l)_L\times SU(r)_R\times U(1), \,r\geq 2; \\
SU(l)_L\times U(1), \,r=1;\\
SU(l)_L, \,r=0.
\end{array}
\right.
\end{equation}

Under the hypothesis adopted, in the given case  of  chiral 
gauge $E_6$  the vacuum condensate in a suitably chosen basis  can  be 
brought to partly diagonal form as follows:
\begin{equation}
<\chi_L\bar \chi_R> =O(\Lambda_X^3)
\underbrace{
\left.\begin{array}{c}
\left(
\begin{array}{ccccc}
&& 0 && \\
\ldots&\ldots&\ldots&\ldots&\ldots\\
1&&&0&\\
&\ddots&&&\\
&&1&&\\
&&&\ddots&\\
&0&&&1
\end{array} 
\right) \\
\\
\end{array}\right.}_{\mbox{\normalsize $r$}}
\begin{array}{ll}
\left.\begin{array}{l}
\\\\
\end{array}\right\}&l-r\\
{\left.\begin{array}{l}
\\
\\
\\
\\
\\
\\
\end{array}\right\}}&r
\end{array}\label{eq2}
\end{equation}
($\Lambda_X$  being the confinement mass scale of the exceptional gauge 
symmetry). This means that all $r$ pieces of $\chi_R$
match some $r$ pieces of $\chi_L$ leaving $n=l-r$ pieces of $\chi_L$ 
unmatched.  

The condensate eq.~\ref{eq2} possesses the following residual symmetry $H$
\begin{equation}
H=\left \{ 
\begin{array}{l}
SU(n)_L\times SU(r)\times U(1), \,r\geq 2;\\
SU(n)_L\times U(1), \,r=1;\\
SU(n)_L,\, r=0,
\end{array}
\right. \label{eq3}
\end{equation}
where $n \equiv l - r$ 
is the net chirality index of the constituents  
$(0 \leq n \leq l)$. 
(Here one should put $SU(n) \equiv  I$ for $n  =  0,$~1.)  In 
the strictly chiral case $(r=0,\, n = l)$ the chiral symmetry is not 
broken at all $(G = H)$, because according to the hypothesis the condensate 
just can not be formed without breaking the $S_{loc}=E_6$  gauge symmetry. 
In other  extreme  vector-like 
case $(l=r,\, n = 0)$ the condensate eq.~\ref{eq2} reduces in terms of  Dirac  
fermions  $\chi =(\chi_L ,\chi _R )$ to the form $<\chi \bar\chi > 
\sim\mbox{diag}(1,\dots,1)$, and  the  chiral 
symmetry is broken down to the vector-like one. In an intermediate 
case $(0 < n < l)$ the chiral symmetry $G$ is  broken  just  partially 
(but to the maximum allowed extent).
It is clear that all  the constituents are in the vector-like representation 
$n\times \underline 1 \oplus \underline r \oplus\bar r$ under the $SU(r)$ 
unbroken subgroup. The same can be shown to be true for the $U(1)$  
residual subgroup. 

Finally, we conclude that in the most general case of the chiral 
gauge $E_6$  the surviving chiral  symmetry  $H$  is  divided  into  two 
parts: strictly chiral and vector-like  ones.  Accordingly,  there 
are two types of constituents  relative  to  $H$:  $n$  massless  Weyl 
fermions and $r$ Dirac ones, the latter having equal  dynamical masses. 
For the present purposes Dirac constituents are  supposed  to 
be intrinsically massless, though they could have some small explicit
mass $m$ ($m \ll \Lambda_X$). 

\paragraph{Massless composite fermions}

Rather general dynamical arguments require that chiral anomalies should 
match at the constituent and composite levels~\cite{thooft,frishman}. 
In this, anomalies for three unbroken  currents  have  to  match  via 
massless composite fermions. Chiral anomaly matching condition  is 
the unique known raison d'~etre  for  the  appearance  of  such  
massless states. Now we  proceed  to  study  this  condition  for  residual 
subgroup $H$ as given by eq.~\ref{eq3}. 

It is well-known~\cite{cvitanovic} that $E_6$  possesses  totally symmetric invariant
tensor $d_{abc}$; $a, b, c = 1,\dots,27$ in the fundamental  representation, 
alongside with the  Levi-Civita  tensor  $\epsilon_{abc\cdots}$,  
and  so  allows formation of both three-particle  and  27-particle  
fermion  bound states. In what follows we  restrict  ourselves  only  
with  three particle  composites.  

In general, for chiral gauge $E_6$  there are  three  ``strata''  of 
composite fermions: pure chiral, mixed chiral and vector-like ones,  
built of two kinds of  $E_6$ constituents,  namely,  strictly  chiral  
and vector-like ones. 
Lorentz couplings of constituents have to be chosen  in 
such a way as to  allow for the formation  of  composite  states  of  the 
required chiralities. A priori, left- and right-handed  components 
of Dirac constituents enter  these  states  in  different  Lorentz 
structures, in particular, those with  derivatives.  Hence,  it  is 
admitted  that  these  constituents,  though   being   potentially 
massive, in  some  chiral  environment  could  not  acquire  their 
dynamical  mass,  so  that  corresponding  composites   are   left 
massless.  (In  this  respect  massless  chiral  fermions  somehow 
resemble composite Goldstone bosons, and for this  reason  one should 
think that potential models are not applicable to them.)

Now, let $\nu (\rho)$ be the chiral index of the state  $\rho$  
(i.e.\  the number of the corresponding 
left-handed composite fermions minus  that  of  the  right-handed  ones).  
$\nu (\rho)$  is  some  unknown  integer  which  is 
supposed to be eventually determined by the  underlying  dynamics. 
The chiral anomaly matching condition  can  just  somewhat  restrict  the 
allowable  sets  of  these  indices.  Note  that  states  composed 
exclusively of Dirac constituents  should have zero indices
due to the discrete LR-symmetry. Appropriate fermions fill in the vector-like  
composite  stratum  and have masses $O(\Lambda_X)$ (\`a la QCD hadrons).

The only triangle anomaly at the level of constituents 
is  that $[SU(n)_L ]^3$  for three $SU(n)_L$  currents. 
Solving the anomaly matching conditions  one obtains in the most general case
a three-parametric set of solutions. These solutions are, in general, not 
vector-like relative to the $SU(r)\times U(1)$ subgroup, 
though the constituents are. Nevertheless appropriate anomalies match and are 
equal 0 in both cases.

Further reduction of the allowed set of indices  could  be 
achieved by imposing some additional physical restrictions. One of 
these is the matching condition of the mixed chiral-gravitational  
anomalies for one $U(1)$ and two gravitational 
currents~\cite{gravanomaly}. This results in the requirement  $\sum Y = 0$, 
where $Y$ is the generator of the $U(1)$ subgroup of the unbroken chiral 
symmetry $H$. This gives one more relation for indices.
The other possible restriction is decoupling  condition~\cite{decoupling} 
(in more refined form, persistent mass  condition~\cite{persist} or 
the  constituent 
number independence~\cite{numberindep}). But for chiral gauge theories the decoupling
conditions is not obligatory~\cite{nodecoupling}.

In addition to massless composite fermions there also appear composite 
Goldstone bosons. They correspond to 
the broken symmetry $G/H$ and are built of one chiral Weyl and one vector-like
Dirac constituents. Goldstone bosons saturate the anomaly matching for the
broken currents from $G/H$.
 
Thus the chiral symmetry breaking pattern is just of the 
type required to embed the SM. But in order to built a particular realistic 
composite model based on this binding mechanism one has to specify a lot of
``subtle'' details, such as the quantum numbers of 
constituents, the intrinsic gauge symmetry, the explicit mass terms etc. 
Presently this can not be done unambiguously. 

Nevertheless the scheme do unambiguously produce the key message for the 
``low energy'' effective theory of unified compositeness. Namely, it should be
a nonlinear model $G/H$ with $G$ and $H$ from the sets of eq.~\ref{eq2} and  
eq.~\ref{eq3}, resp. This could be a  starting point for  studying the unified
compositeness at the subthreshold energies. Additional topics of the scheme 
can be found in ref.~\cite{pirogov1}.

\section{Higgs doublet as composite Goldstone boson} 

\paragraph{Nonlinear Standard Model} 

To describe  the ``low energy'' (i.e., below the compositeness scale) 
behaviour of the composite leptons, quarks and Higgs bosons, without detailed 
knowledge of the hyperstrong interactions responsible for their internal 
substructure, one has to refer to the framework of the effective  
field theory~\cite{weinberg1}. In essence, it requires just the assumption 
about the symmetry breaking pattern $G\to H$, as well as
the light particle content. The simplest nonlinear model $G/H$
to implement the idea of the Higgs doublet as composite Goldstone boson 
was first proposed on  phenomenological grounds in ref.~\cite{georgi}. 
It was further refined from the unified compositeness point of view 
and systematically studied in ref.~\cite{pirogov2}. It may be called 
the minimal Nonlinear Standard Model (NSM). In what follows we present the  
basic features of the NSM. More details can be found in ref.~\cite{pirogov2}.

It can be shown that the simplest nonlinear model $G/H$ with the required 
properties is based on the symmetry breaking pattern $G=SU(3)\times U(1)$ and
$H=I_{SM}=SU(2)_L\times U(1)_Y$. The extended symmetry $G$ contains the broken 
isodoublet generators $X_I$, $X^{\dagger I}$,  $I=1,2$, as well as the broken 
hypercharge $Y'$ in addition to the unbroken SM generators   
of the weak isospin $T_i$, $i=1,2,3$ and the weak hypercharge $Y$.
With the broken generators of $G/H$ there are associated the Goldstone doublet 
$\phi_I$ and singlet $\phi'$. This extra Goldstone boson is absorbed by
the gauge boson of the additional dynamically broken local symmetry 
$U(1)_{Y'}$. 
The latter is the minimum one required to eventually convert 
the true Goldstone doublet $\phi$ via the radiative corrections 
into the SM Higgs doublet. In this prototype model, the QCD colour symmetry 
is supposed to be trivially present on both sides of the symmetry breaking 
chain.

\paragraph{Nonlinear realization} 

As the nonlinear model $G/H$, the NSM can be built via the canonical nonlinear 
realization of the symmetry $G$ that becomes linear when restricted to 
$H$~\cite{callan}. The Goldstone bosons parameterize  the element of 
the left coset space
$\xi\in G/H$
\begin{equation}
\xi=e^{i\phi'Y'/{\cal F}'}e^{i(\phi_I X^{\dagger I}+{\mathrm h.c.})/{\cal F}},
\end{equation}
with ${\cal F}$, ${\cal F}'={\cal O}({\cal F)}$ being the symmetry breaking 
mass scales.
Here $\xi$ and $\phi$ transform under $g\in G$ as
\begin{eqnarray}
g&:&\xi\,\,\to\,\,\tilde \xi=g\xi h^\dagger(g,\xi),\nonumber\\
&&\phi\,\,\to\,\,\tilde\phi(g,\xi),
\end{eqnarray}
where $h(g,\xi)$ and $\tilde\phi(g,\xi)$ are uniquely determined through
the natural decomposition
\begin{equation}
g\xi\equiv\tilde \xi h=e^{i\tilde\phi'Y'/{\cal F}'}e^{i(\tilde\phi_I
X^{\dagger I}+{\mathrm h.c.})/
{\cal F}}h.
\end{equation}
A matter field $\psi$ transforms under $g\in G$ as
$$g\,:\,\,\psi\,\,\to \rho ( h(g,\xi))\psi,$$
where $\rho$ is a linear representation of $H$, and $h(g,\xi) $ is determined
by the equation above.

Derivatives of the Goldstone and matter fields enter through the 
Maurer-Cartan 1-form
$\Delta_\mu\equiv 1/i\, \xi^\dagger\! D_\mu \xi $, with
$D_\mu$ being the  derivative covariant v.r.t.\  the gauge symmetry $I_{loc}$.
The 1-form $\Delta_\mu$
contains the nonlinear covariant derivative ${\cal D}_\mu\phi$ of the
Higgs-Goldstone doublet $\phi$, as well as a part required to construct the
nonlinear covariant derivative ${\cal D}_\mu \psi$ of the matter fields $\psi$.
Namely, let us divide $\Delta_\mu$ into two parts: $\Delta_{\|\mu}$ which is
parallel to $G/H$  and $\Delta_{\perp\mu}$ orthogonal to it, along the
unbroken symmetry $H$: 
\begin{eqnarray}
\Delta_\mu&=&(\Delta_{\|I\mu}X^{\dagger I}+{\mathrm h.c.})+
\Delta^{0\prime}_{\|\mu}Y'
\nonumber\\ \label{Delta}
&+&\,\Delta^i_{\perp\mu}T^i+\Delta^0_{\perp\mu}Y.
\end{eqnarray}
Then one has
\begin{eqnarray}
({\cal D}_\mu\phi)_I/{\cal F}&=&\Delta_{\|I\mu},\nonumber\\
{\cal D}_\mu\psi&=&\Bigl(\partial_\mu+i(\Delta^i_{\perp\mu}T^i+\Delta^0_{\perp
\mu}Y)\Bigr)\psi.
\end{eqnarray}
 
All the terms in eq.~\ref{Delta} transform nonlinearly under $G$ as the
irreducible representations of $H$  and can be used to construct 
the effective Lagrangian of the NSM. It consists of the most general
superficially $H$ invariant expressions built of the $\psi$'s 
(but not of $\phi$'s) and the nonlinear covariant derivatives
${\cal D}_\mu\phi$ and ${\cal D}_\mu\psi$, the latter ones transforming
like $\psi$'s under the nonlinearly realized extended symmetry~$G$.
Additional building blocks are given by
the nonlinear generalization of the gauge field strengths 
$\Delta_{\mu\nu}\equiv 1/i\, \xi^\dagger [D_\mu,D_\nu] \xi $.

Let us decompose ${\cal L}_{eff}$ into the gauge, Higgs and fermion parts:
\begin{equation}
{\cal L}_{eff}={\cal L}_G+{\cal L}_H+{\cal L}_F.
\end{equation}
Then the gauge part ${\cal L}_G$ of the Lagrangian is built of 
the irreducible under $H$ components of $\Delta_{\mu\nu}$, the latter ones  
being defined as in eq.~\ref{Delta}. The Higgs part
\begin{equation}
{\cal L}_H=({\cal D}_\mu\phi)^\dagger({\cal D}_\mu\phi)+{\cal O}(1/{\cal F}^4)
\end{equation}
is uniquely determined by the symmetry breaking pattern. 
And finally, the fermion part for the 
left-chiral fermion fields $\psi_\rho$, belonging to the irreducible 
representation $\rho$ of 
the unbroken subgroup $H$, takes the form
\begin{eqnarray} \label{nsm1}
{\cal L}_F&=&\sum_\rho(m_\rho\psi_{\bar \rho}\psi_\rho+{\mathrm h.c.})+
\sum_\rho\bar\psi_\rho
\sigma_\mu
i/2 \stackrel{\leftrightarrow}{{\cal D}}_\mu\psi_\rho\nonumber\\
&+&\sum_\rho\eta_\rho\bar\psi'_\rho\sigma_\mu\psi_\rho\Delta^{0\prime}_{\|\mu}+
{\mathrm h.c.}\nonumber\\
&+&\frac{1}{{\cal F}}\sum_{\rho_1\rho_2}\chi_{\rho_1\rho_2}\bar\psi^I_{\rho_2}
\sigma_\mu
\psi_{\rho_1}({\cal D}_\mu\phi)_I+{\mathrm h.c.}\nonumber\\
&+&\frac{1}{{\cal F}}\sum_{\rho_1\rho_2}\bar\chi_{\rho_1\rho_2}\bar
\psi^I_{\rho_2}
\sigma_\mu\psi'_{\rho_1}\overline{({\cal D}_\mu\phi)}_I+{\mathrm h.c.}+
{\cal O}(\frac{1}{\cal F}),
\end{eqnarray}
where $\overline{({\cal D}_\mu\phi)}_I\equiv(i\tau_2)_{IJ}({\cal D}_\mu\phi)^{\dagger J}$.
We have omitted terms irrelevant for the later discussion.
Here $m$, $\eta$, $\chi$ and $\bar\chi$ are arbitrary parameters. 
The first expression in ${\cal L}_F$ describes the explicit mass terms of the 
vector-like heavy composite fermions ($m_\rho={\cal O}({\cal F})$).
It can be shown that all the terms ${\cal O}(1/{\cal F})$ mix with necessity 
the light chiral and heavy vector-like fermions.
In the limit ${\cal F}\to\infty$ the finite part of ${\cal L}_{eff}$ reproduces
exactly the SM  Lagrangian (except for the Higgs potential and Yukawa 
interactions). In this limit the heavy vector-like fermions decouple 
from the SM light sector.

\paragraph{Higgs and Yukawa interactions}
In reality, symmetry $G$ is not exact but explicitly violated, e.g., 
by the extended electroweak interactions, since only part of $G$, 
namely $I_{loc}=SU(2)_L\times U(1)_Y\times U(1)_{Y'}$, is supposed to be gauge.
Gauge radiative corrections may lead to a misalignment of the 
dynamically unbroken subgroup $H$ relative to the gauge $I_{loc}^{SM}$.
This results in the spontaneous SM symmetry breaking and the appearance of 
the Higgs and Yukawa effective interactions. 
This effect may be properly accounted for by adding  the symmetry violating
effective Lagrangian  
\begin{equation}
\Delta{\cal L}_{H}={\cal F}^4\biggl(\bar g^2\mbox{tr}(\xi^\dagger T_i\xi T_i)+\bar g^2_1\mbox{tr}(\xi^
\dagger Y\xi Y)-\bar g'^2_1\mbox{tr}(\xi^\dagger Y'\xi Y')\biggr). \label{Higgs}
\end{equation}
Here the effective couplings $\bar g^2$, $\bar g^2_1$ and $\bar g'^2_1$ are 
equal
to the product of the corresponding gauge constants squared and some spectral 
integrals. Note the difference in the sign between  the contributions of the 
dynamically broken and unbroken gauge interactions~\cite{dashen,peskin2}.
Decomposition of $V_{H}=-\Delta{\cal L}_{H}$ in the region of weak fields 
($|\phi|/{\cal F}\ll 1$) gives the Higgs potential up to 
${\cal O}(1/{\cal F}^2)$. Note that the Higgs boson is expected naturally 
to be light in the scheme.

As for the Yukawa interactions, ${\cal L}_{eff}$ includes three 
ingredients required for their appearance: the chirality changing 
mass terms of the heavy vector-like fermions ($m={\cal O}({\cal F})$), 
the Goldstone interactions of this fermions 
($\sim {\cal D}_\mu\phi/{\cal F}$), and, finally, the weak gauge mixing of 
the light chiral and heavy vector-like fermions. So, the loop corrections 
may lead to the appearance 
of the symmetry violating effective Lagrangian like eq.~\ref{Higgs}.
Its decomposition can be shown to result in the nonderivative Yukawa 
couplings of order ${\cal O}(g^2/(4\pi)^2)$. More details can be found
in ref.~\cite{pirogov2}.

\section{Vector boson dominance of gauge interactions}

\paragraph{Hidden local symmetry}

Being a nonlinear model $G/H$, the NSM
is equivalent to the model with linearly realized symmetry $G\times
\hat H_{loc}$~\cite{kugo}. Here $\hat H_{loc}\simeq H$ is the hidden local 
symmetry of the original NSM with the appropriate auxiliary
gauge bosons. In the context of the minimal NSM the phenomenon of the hidden 
local symmetry has been first studied in ref.~\cite{pirogov3}. The essence of 
the latter one is as follows.

In the linear model, the field variable is the element of the whole 
group $G$ which can be parameterized as
\begin{equation}
\hat \xi=\xi h,\,\,\,h\in H.\label{hatxi}
\end{equation}
The following transformation law under $g\times\hat h(x)\in G\times 
\hat H_{loc}$ takes place:
\begin{equation}
g\times \hat h(x):\,\,\hat \xi\to g\hat \xi\hat h^\dagger(x).
\end{equation}
The linear model describes dynamical/spontaneous symmetry breaking
$G\times \hat H_{loc}\,\to\,H$, with the total local symmetry being broken as 
$I_{loc}\times\hat H_{loc}
\,\to\,I_{loc}^{SM}=SU(2)_L\times U(1)_Y$.

To construct the Lagrangian of the linear model one has to introduce
the modified 1-form $\hat\Delta_\mu=1/i\,{\hat \xi}^\dagger 
\hat D_\mu\hat \xi$, with
$\hat D_\mu$ being now the derivative covariant both under the intrinsic 
gauge symmetry $I_{loc}$ and the hidden local symmetry $\hat H_{loc}$.
Let us again divide $\hat\Delta_\mu$ into two parts: $\hat\Delta_{\|\mu}$ 
and $\hat\Delta_{\perp\mu}$.
Under $G\times\hat H_{loc}$ the longitudinal  part 
$\hat\Delta_{\|\mu}$ transforms homogeneously as in the original 
nonlinear model, and so does now the transversal part $\hat\Delta_{\perp\mu}$. 
It is precisely the auxiliary vector fields $\hat W_\mu^i$ and 
$\hat S_\mu$, corresponding to $\hat H_{loc}$ which make the
transformation of $\hat\Delta_\perp$ homogeneous. In the unitary under
$\hat H_{loc}$ gauge, i.e.\ at $h\equiv 1$ in eq.~\ref{hatxi}, the modified 
1-form looks like 
\begin{eqnarray}
\hat \Delta_{\|\mu}&=&\Delta_{\|\mu},\nonumber\\
\hat \Delta_{\perp\mu}^i&=& \Delta^i_{\perp\mu}-\hat g\hat W^i_\mu,\\
\hat \Delta^0_{\perp\mu}&=&\Delta^0_{\perp\mu}-\hat g_1\hat S_\mu,
\nonumber
\end{eqnarray}
where $\Delta_\mu$ is the 1-form present in the original  minimal NSM,
 $\hat g$ and $\hat g_1$ being some new strong coupling constants (supposedly,
$\hat g^2/4\pi ={\cal O}(1)$).
 
In the effective Lagrangian of the linear model, the new terms appear.
They are related with the orthogonal part of the modified 1-form.
Here are some of the appropriate terms in the gauge sector:
\begin{equation}
{\cal L}_G=\frac{\lambda{\cal F}^2}{2}(\hat\Delta_{\perp\mu}^i)^2+\frac{\lambda_1
{\cal F}^2}{2}(\hat\Delta_{\perp\mu}^0)^2+\cdots,
\end{equation}
and for the chiral fermions they are
\begin{eqnarray}
{\cal L}_F&=&\bar\psi\sigma_\mu i(\partial_\mu+i\hat g\hat W^i_\mu T^i+i\hat g_1\hat S_\mu
)\psi\nonumber\\
&+&\kappa\bar\psi\sigma_\mu T^i\psi\hat\Delta_{\perp\mu}^i+\kappa_1\bar\psi
\sigma_\mu Y\psi\hat\Delta_{\perp\mu}^0+\cdots.
\end{eqnarray}
Here $\lambda$'s and $\kappa$'s are free parameters. It is to be noted that the
matter fields $\psi$ transform now only under $\hat H_{loc}$. The modified
covariant derivative  for them contains only the composite
$\hat W_\mu$ and $\hat S_\mu$, but not the elementary $W_\mu$ and $S_\mu$,
the latter ones entering only through the nonminimal interactions.

Introducing the vector fields in such a way  without kinetic terms 
is just a formal procedure. But we believe that the required 
kinetic terms are developed
by the quantum effects, and the new composite vector bosons become physical. 
This takes place, e.g., in 2- and 3-dimensional nonlinear 
$\sigma$-models~\cite{dadda},
as well as in the hadron physics as accomplished fact.

\paragraph{Vector boson dominance}

From the Lagrangian of the linear model, one can read off the Lagrangian 
terms of the vector boson-current interactions:
\begin{eqnarray}
{\cal L}_{int}&=&-gW_\mu^i\Bigl((1-\lambda)J_\mu^i(\phi)
+\kappa J_\mu^i(\psi)\Bigr)
\nonumber\\
&&-\hat g \hat W_\mu^i\Bigl(\lambda J_\mu^i(\phi)+(1-\kappa)J_\mu^i(\psi)
\Bigr).  
\end{eqnarray}
Here $J_\mu^i(\psi)=\bar\psi\gamma_\mu T^i\psi$ and $J_\mu^i(\phi)=
\phi^\dagger i\tau^i/2\!\stackrel{\leftrightarrow}{D}_\mu\!\phi$ are the 
usual SM isotriplet currents, with $D_\mu$ being the
SM covariant derivative. To these isospin terms, one has to add the similar 
hypercharge isosinglet terms.
Impose now the natural requirement that all the composite particles $\phi$ and 
$\psi$ 
interact directly only with the composite vector bosons $\hat W$ and $\hat S$,
but not with the elementary ones $W$ and $S$. In  other words, this is
the well-known hypothesis of the vector boson dominance (VBD).
This requirement allows one to fix the free parameters: $\lambda=1$,
$\kappa=0$ and similarly for the isosinglet parameters.

The terms $(\hat\Delta^i_\perp)^2$ and $(\hat\Delta^0_\perp)^2$ 
describe the mass mixing of the elementary and composite gauge bosons,
namely,
$W$ with $\hat W$ and $S$ with $\hat S$. Diagonalizing these terms one gets 
two sets of physical vector bosons: the massless isotriplet and isosinglet 
physical bosons $\bar W^i$ 
and $\bar S$, as well as the massive ones $\bar{\hat W}^i$ and $\bar {\hat S}$
with masses of order ${\cal F}$.
Due to the  heavy physical vector boson exchange, the new low energy effective
current-current interactions appear in addition to that of the SM:
\begin{eqnarray}
{\cal L}_{int}^{(VBD)}&=&-\frac{1}{2{\cal F}^2}\Bigl(J^i_\mu(\psi)J^i_\mu(\psi)
+\eta_1J^0_\mu(\psi)J^0_\mu(\psi)\Bigr)\nonumber\\
&&-\frac{1}{{\cal F}^2}\Bigl(J^i_\mu(\psi)J^i_\mu(\phi)+\eta_1J^0_\mu(\psi)
J^0_\mu(\phi)\Bigr).
\end{eqnarray}
Here $\eta_1$ is a free parameter, related to the original  minimal NSM.
Note that the VBD does not affect the low energy Higgs boson 
self-interactions, the latter ones being 
determined by the original minimal NSM alone:
\begin{eqnarray}
{\cal L}_{int}(\phi)=-\frac{1}{{\cal F}^2}\Bigl(\frac{1}{3}J^i_\mu(\phi)
J^i_\mu(\phi)+J^0_\mu(\phi)J^0_\mu(\phi)\Bigr),
\end{eqnarray}
which could in principle be simplified by the Fiertz rearrangement.
All these expressions are valid only at energies $\sqrt{s}\ll{\cal F}$.

To resume, the unified compositeness plus the VBD prescribe the 
two-parameter set of the universal residual 
fermion-fermion, fermion-boson and boson-boson  interactions, with their 
space-time and internal structure being fixed including sign.
The unified compositeness scale ${\cal F}$ is expected to be in the deca-TeV region.
Hence, the TeV energies are required to probe these new contact interactions.

\section{Universal dominant residual interactions}

\paragraph{VBD of electroweak interactions}

We have investigated the potential to test the hypothesis of the VBD of
electroweak interaction at the future 2~TeV $e^+e^-$  linear collider via 
$e^+e^-\to \bar ff$~\cite{kabachenko1} and $e^+e^-\to ZH$, 
$W^+W^-$~\cite{kabachenko2}. We chose for studying  a set of integral 
characteristics: the relative deviation $\Delta$ in the total cross-sections 
from the SM values,
the forward-backward charge asymmetry $A_{FB}$,
the left-right polarization asymmetry $A_{LR}$
and the mixed asymmetry $A_{LR}^{FB}$.

We have calculated these observables for the processes 
$e^+e^-\to \mu^+\mu^-\,(\tau^+
\tau^-)$, $\bar bb$, $\bar cc$, $jet\,\,jet$ and for the Bhabha scattering
$e^+e^-\to e^+e^-$ as functions of the parameter $\eta_1$ for the 
various values 
of ${\cal F}$. The general results of these calculations are as follows.
For all the processes (except Bhabha scattering) all the 
asymmetries have the similar behaviour. 
First of all, there exists a particular value
of $\eta_1=\tan^2\theta_W\simeq0.3$ when all the asymmetries coincide with 
those of the SM. The only way to unravel the contact interactions in this 
particular case is to study directly the total cross-sections.
Another particular value of $\eta_1=g_1^2{\cal F}^2/s$ provides the best 
case for studying the contact interactions, when all the asymmetries in all 
the processes saturate their maximal values.

To evaluate the statistical significance of the observed deviations we 
have considered the 
total cross-sections. Fig.~1 presents the reach for the scale ${\cal F}$ at
$2\sigma$ level (95\% C.L.) via the total cross-sections in the 
various $\bar ff$ channels. 
To this end we took into account only the statistical
errors and accepted the integrated luminosity $\int {\cal L}\,dt$
moderately to be 20$\,\,fb^{-1}$. 
In the case of the Bhabha scattering $e^+e^-\to e^+e^-$ an optimal value of the
cutoff, equal to 0.85, was chosen. Here the sensitivity is maximal due to 
the maximal
suppression of the $t$-channel peak at the statistics still high enough. It is
seen that in the processes $e^+e^-\to \bar ff$ the VBD can be tested 
for the unified substructure scale ${\cal F}$ up to ${\cal O}(50\,\,$TeV).

For the processes $e^+e^-\to ZH$ and $W^+W^-$, it proved to be of
importance to consider the
polarized cross-sections $\sigma(P_e)$, with
$P_e$ denoting the polarization of electron beam (the positron beam was
taken to be unpolarized). So, we have studied the relative deviation 
$\Delta(P_e)$ in the 
polarized cross-section from that of the SM.
In  the cases of both $ZH$ and $WW$ pair production one has
$|\Delta(-1)|\ll|\Delta(+1)|$. Hence 
one is lead to  conclude that it is preferable to operate with the maximum 
right-handedly
polarized electrons to observe as large deviations in the total cross-sections
from the SM values as possible. The advantage
of the right-handed polarization can  be seen, e.g., from the picture that
presents the  scale ${\cal F}$ versus the parameter
$\eta_1$, attainable at 95\% C.L.\ (Fig.~2).

Thus, using the right-handed polarized electron beam the VBD can be tested up 
to the scale ${\cal F}$ of the order of 25 TeV in the $e^+e^-$ annihilation into
boson pairs. Here the calculations for the $W^+W^-$ 
pair production have been made under
the instrumental cutoff $|\cos\theta|\leq0.8$. In addition, an
optimal cutoff in the forward direction, whose sense is similar to that
in the forward Bhabha scattering, has been found to be
$\cos\theta\leq 0.3$.

\paragraph{Anomalous triple gauge interactions}

In addition to the  VBD interactions, a lot of other ``low energy'' residual
interactions is allowed in the scheme of the unified compositeness.
 In particular, the exotic triple gauge interactions (TGI)~\cite{hagiwara} 
are conceivable too, and can contribute to the
$W^+W^-$ pair production. The question arises as to what extent the two
types of new interactions could imitate each other. 

The anomalous TGI should  originate from a kind of the SM extension.
Here, the  SM symmetry $SU(2)_L\times U(1)_Y$
could be realized either linearly or nonlinearly.
In the case of the nonlinear realization (being still linear on the $U(1)_{em}$
subgroup), the nonlinearity scale $\Lambda$ is just the SM v.e.v.\ $v$.
Thus, this kind of extension has nothing to do with the unified compositeness
we consider. On the other hand, for the linear SM symmetry realization 
the scale $\Lambda$ is not
directly related with $v$ and could be as high as desired. Thus, we chose 
it to be the unified compositeness  scale ${\cal F}={\cal O}(10\,\,$TeV).

All the conceivable linearly realized residual
interactions are described by the $SU(2)_L\times U(1)_Y$ invariant operators
built of the SM fields~\cite{leung,derujula}. All the operators
which are relevant to the anomalous TGI vertices are naturally expected
to be  ${\cal O}(g)$ or less in the gauge couplings, but one exception, 
namely, ${\cal O}_{WS}$. The latter stems from the nonlinear generalization of 
the field strengths in the NSM. The similar gauge kinetic terms of the
isotriplet $W$ and isosinglet $S$ bosons have no  gauge couplings.
So, the same must naturally happen for ${\cal O}_{WS}$, for its origin is
of the same nature.

Thus, we have retained the ${\cal O}_{WS}$ operator alone and have chosen
the proper effective Lagrangian to be
\begin{equation}
{\cal L}_{eff}=\frac{C}{2}\frac{1}{{\cal F}^2}{\cal O}_{WS}
\equiv\frac{C}{2}\frac{1}{{\cal F}^2}
\phi^\dagger\frac{\tau_i}{2}\phi
W^i_{\mu\nu}S_{\mu\nu},
\end{equation}
where $C={\cal O}(1)$.
With account for all the contributions from this operator we have found
that the deviations from the SM predictions even in the most enhanced
TGI case are much smaller then those in the VBD case.   
So, the VBD is in fact dominant.

\section*{Conclusion} 

The scenario of unified compositeness of leptons, quarks and Higgs bosons,
with the unification of the Higgs and Yukawa interactions as residual ones, 
is the viable alternative to presently popular scenarios of New Physics 
with the elementary point-like fields and fundamental interactions. 
This scenario 
allows one to have a fresh look at the old problems and to put forward the 
new ones. The naturally preferred Deca-TeV compositeness scale makes the 
scenario amenable to experimental study at the future TeV energy colliders.  
If realized in Nature, this scenario would open completely new perspectives 
for the whole high energy physics development.

\paragraph{Acknowledgements}
This work is supported partially by the RFBR under grant No.~96-02-18122a 
and partially by the Competition Center for Fundamental Natural Sciences 
under grant No.~95-0-6.4-21.

\newpage

\thispagestyle{empty}
{\Large\bf Figure Captions}
\vspace{2cm}

{\bf Fig. 1}: The reach at 95\% C.L.\ for the compositeness scale ${\cal F}$,
vs.\ the parameter $\eta_1$, via studying the total cross-sections of the
processes $e^+e^-\to \bar ff$.
\vspace{2cm}

{\bf Fig. 2}: The same as in Fig.~1 for the processes  $e^+e^-\to ZH$, 
$W^+W^-$ with the various electron polarizations $P_e$ ($m_H=200\,\,$GeV).

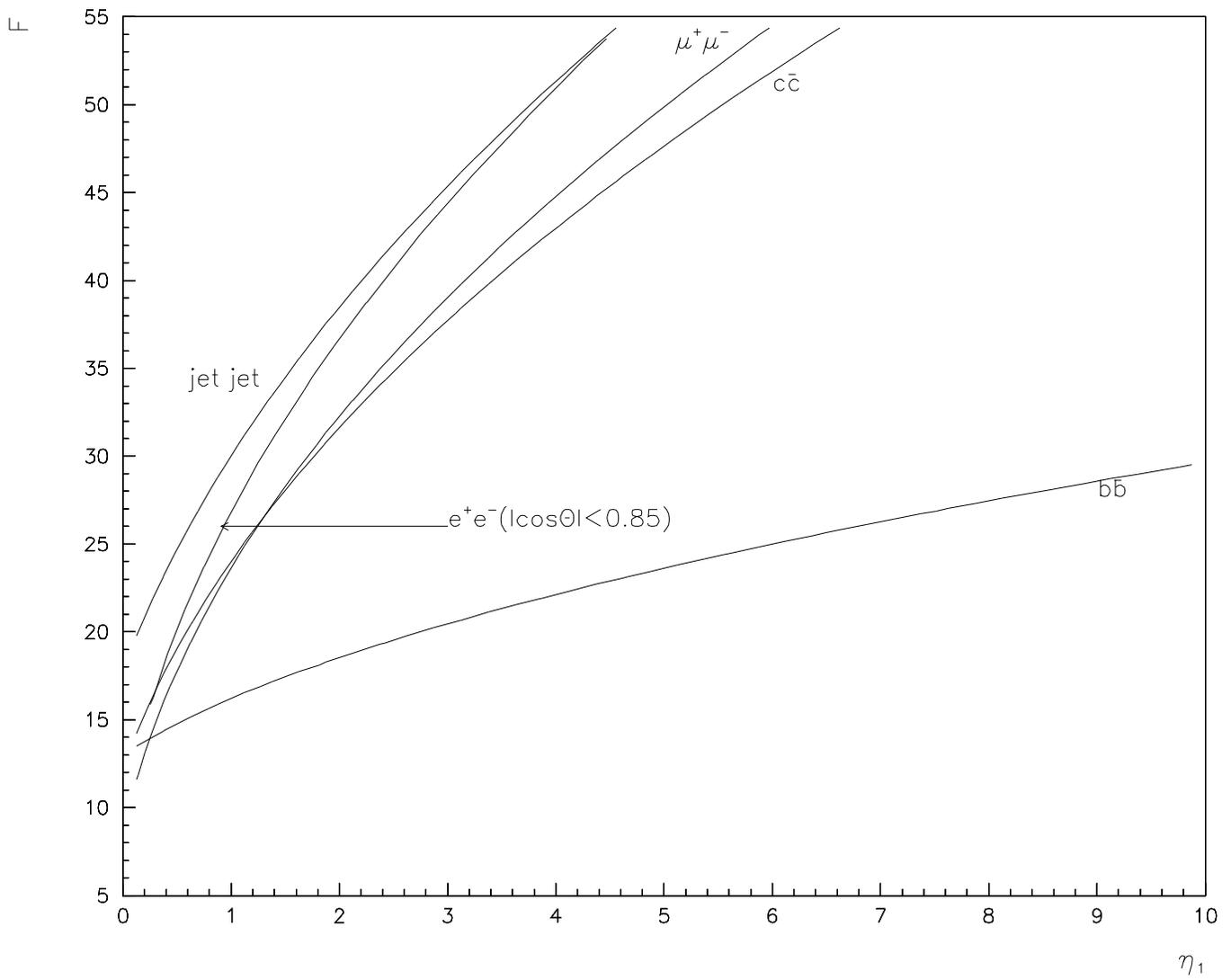
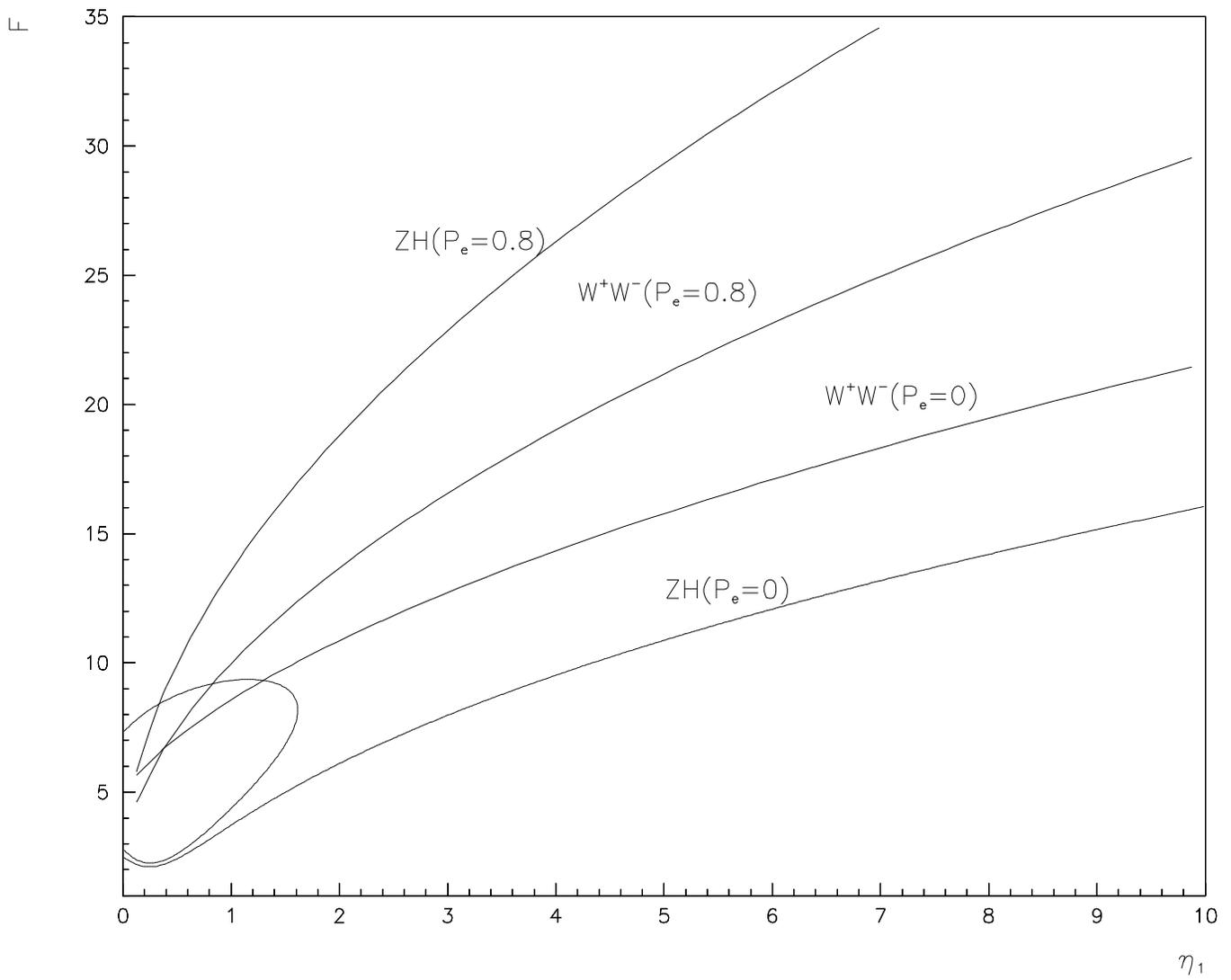
\end{document}